\documentclass[reprint,amsmath,amssymb,braket,aps,prb,longbibliography]{revtex4-2}
\usepackage{graphicx,color,braket}
\usepackage{dcolumn}
\usepackage{bm}
\usepackage[colorlinks=true,linkcolor=blue]{hyperref}
\usepackage{colortbl}

\begin{document}
	
	\title{Electromagnetic drag in partly gated 2d electron system via highly confined screened plasmons}

	\author{I.M. Moiseenko}
	\affiliation{Center for Photonics and 2d Materials, Moscow Institute of Physics and Technology, Dolgoprudny 141700, Russia}
    \email{moiseenko.im@mipt.ru}
    
    \author{D.A. Svintsov}
	\affiliation{Center for Photonics and 2d Materials, Moscow Institute of Physics and Technology, Dolgoprudny 141700, Russia}
   \email{svintcov.da@mipt.ru}

    \author{Zh.A. Devizorova}
	\affiliation{Center for Photonics and 2d Materials, Moscow Institute of Physics and Technology, Dolgoprudny 141700, Russia}

\begin{abstract}
Generation of photocurrent via photon drag effect enables very fast light detection with response time limited by momentum relaxation. At the same time, photon drag in bulk uniform samples is small by the virtue of small photon momentum. We show that the edge of metal gate placed above a two-dimensional electron system (2DES) provides highly non-uniform electromagnetic field that enhances the drag effect. We study the drag photovoltage using an exact solution of diffraction problem for 2DES with semi-infinite metal gate. We show that the only non-trivial dimensionless parameters governing the drag responsivity are the 2DES conductivity scaled by the free-space impedance $\eta$ and gate-2DES separation scaled by the incident wavelength $d/\lambda_0$. For radiation with electric field polarized orthogonal to the gate edge, the responsivity is maximized for inductive 2d conductivity with ${\rm Im}\eta \sim 1$ and ${\rm Re}\eta \ll 1$, and becomes very small for the capacitive 2d conductivity. The electromagnetic ponderomotive force pushes the charge carriers under the gate at arbitrary 2d conductivity, and the force direction is opposite to that at metal-2DES lateral contact. These patterns are explained by the dominant role of gated 2d plasmons in the formation of PDE photovoltage.
\end{abstract}
	
\maketitle
\section{Introduction}

Studies of photocurrent generation in low-dimensional systems can shed light on rich microscopic physical processes~\cite{Koppens_diagnostic_photocurrent}, including carrier scattering~\cite{Kachorovskii_Hydrodynamic,Beklov_tellurium}, thermalization~\cite{Herrero_cooling}, recombination~\cite{Rumyantsev_PC_lifetime} and, in selected cases, on the symmetry and topology of their wave functions~\cite{Koppens_MATBG_photocurrent,Delgado-Notario2025}. Such studies are generally complicated by deep-subwavelength size of most low-dimensional structures, abundance of metal contacts, gates, and irregular sample shapes~\cite{Candussio2020,Bandurin2021}. Understanding the general laws of photocurrent generation at basic elements of low-dimensional structures is important for correct understanding of such fundamental experiments. At the same time, such understanding is necessary for optimization of electromagnetic detectors based on low-dimensional systems, which have already achieved very good performance metrics~\cite{Titova2024,Castilla_fast_THz}.

An edge of the metal gate placed above a two-dimensional system (2DES) represents one of the main building blocks where the photocurrent generation occurs. If the gate voltage is applied, the non-uniformity of carrier density acts as a radiation rectifier, which was proved by measurements of local photovoltage~\cite{Muravev_detector,Gabor_thermoelectric}. Even in the absence of gate voltage, the photocurrent generation can still occur due to the non-uniformity of the local field. Such non-uniformity leads to the ponderomotive drag force on charge carriers and direct momentum transfer from electromagnetic field~\cite{Durned_structured,Svintsov_drag_contact}. The effect can be considered as a sub-wavelength analogue of the photon drag. In some papers, it is referred to as 'plasmonic drag' to underline its sub-wavelength nature~\cite{Fateev_drag_asymmetric_graphene,Popov_noncentrosymmetric,Popov_plasmonic_drag}. For symmetrically placed metal gate, the ponderomotive forces at its opposite edges compensate each other, thus the total photocurrent in 2DES is zero. In asymmetric structures, particularly with overlaying gate and electric contact, the compensation generally does not occur, and the resulting photovoltage is measurable.

Our paper addresses the photon drag effect (PDE) in partially gated 2DES and derives its functional dependences on structure parameters, including carrier density, 2D conductivity, electromagnetic frequency and gate-channel separation. Our main finding is large PDE in partly gated setting, not limited by the small in-plane momentum of photon, and emerging already for the normal light incidence. Surprisingly, the direction of ponderomotive force (and photocurrent) in partly gated setting is opposite to that for lateral metal-2DES contact. Even for very small gate-2DES separation, the gate edge efficiently excites gated (screened) 2d plasmons. They appear dominant for generation of PDE photovoltage, even in the overdamped regime where the 2DES conductivity is purely real. All the more, the PDE becomes very small for capacitive 2D conductivity that does not allow the excitation of transverse magnetic plasmons.

Differently from preceding numerical~\cite{Fateev_drag_asymmetric_graphene,Popov_noncentrosymmetric,Ludwig_PTE_vs_RSM} and approximate analytical~\cite{Detection_mixing,Sakowicz2011} approaches, we exploit here an {\it exact} solution for electromagnetic scattering at the partially gated 2DES~\cite{Moiseenko_diffraction_launching}. The exact spatial Fourier spectrum of the diffracted field $E(q)$ is then used to build the PDE photovoltage developed at the 2DES. Naturally, the photovoltage $V_{\rm ph}$ appears proportional to the degree of field asymmetry $q[|E(q)|^2 - |E(-q)|^2]$ as a function of wave vector $q$, integrated over all wave vectors~\cite{Popov_noncentrosymmetric}. Use of an exact solution allows us to single out the functional dependences of PDE on system parameters. Namely, $V_{\rm ph}$ is inversely proportional to the carrier density $n$ and electromagnetic frequency $\omega$, and directly proportional to the dimensionless momentum transfer coefficient $\alpha$. The latter is parametrized by only two dimensionless parameters: the dimensionless 2DES conductivity $\eta = 2\pi \sigma/c$ (in Gaussian units) and gate-channel separation in units of incident wavelength $2\pi d/\lambda_0 \equiv k_0 d$. The momentum transfer $\alpha$ reaches its maximum for 2DES with weak dissipation ${\rm Re}\eta \ll 1$ and moderately large inductance ${\rm Im}\eta \sim 1$. These parameters correspond to the favourable conditions for plasmon launching.

\section{Electromagnetic drag in partly gated 2DES}

\subsection{Exact solution for the local field in partially gated 2DES}

\begin{figure}[ht!]
\centering
\includegraphics[width=0.9\linewidth]{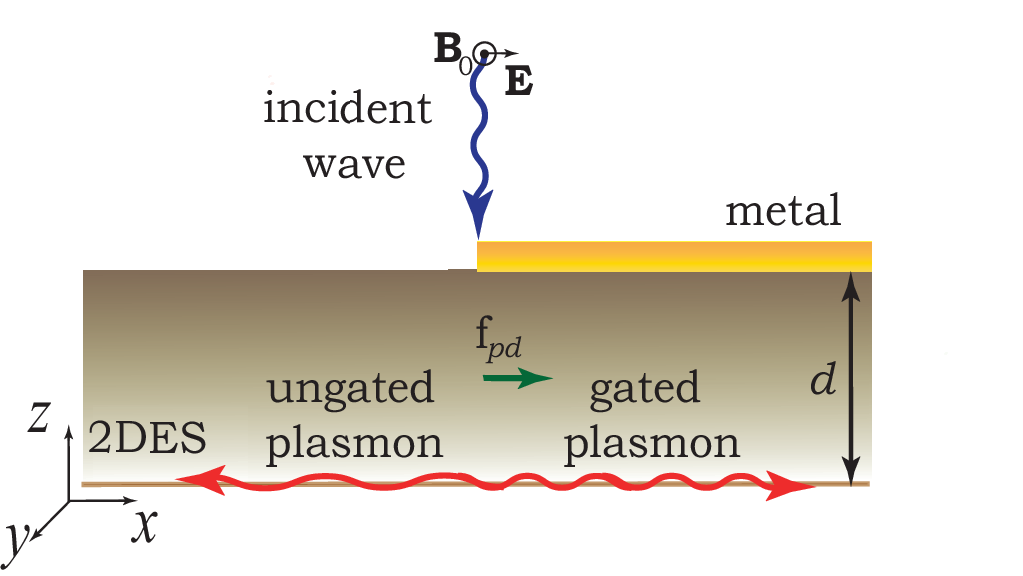}
\caption{Schematic of the studied structure: a partly gated 2DES illuminated by a normally incident plane wave with electric field orthogonal to the edge. Diffraction of the incident wave results in efficient launching of both gated and ungated plasmons (least for inductive 2DES conductivity), which produce drag force $f_{pd}$ on the charge carriers.}
\label{fig:Structure}
\end{figure}

The structure under study is shown in Fig.~\ref{fig:Structure} and represents a 2DES covered by a semi-infinite gate. The electromagnetic field with frequency $\omega$ and wave number $k_0=\omega/c$ is incident normally onto the structure, the electric vector of the wave is polarized normally to the gate edge. Diffraction of the wave on gate edge leads to local modifications of the field strength, such that the resulting field is highly non-uniform and asymmetric. The latter fact enables the non-zero average drag force. In particular case of clean 2DES with inductive impedance, the field can be represented as a combination of gated and ungated plasmons launched away from the edge. In real experiments, the 2DES should have contacts at the left and right edges, while the gate should have finite size. Neglect of the diffraction at these objects is possible if the illuminating beam has finite extent, i.e. narrower than total 2DES length and gate length. Still, the beam should be wider than the wavelength, such that the effects of non-uniform illumination could be neglected.

An exact solution for electromagnetic diffraction in the partially gated 2DES was obtained using the Wiener-Hopf method in [\onlinecite{Moiseenko_diffraction_launching}] for arbitrary incidence angle. We repeat it here for the normal incidence ($k_x=0$) to focus at the photon drag due to field inhomogeneities. Denoting the incident field amplitude as $E_0$, the Fourier spectrum of the local field in the 2DES plane ($z=0$) becomes 
\begin{multline}
\label{eq-Eq_2d}
E_{\rm{q}}=\frac{iE_{0}}{1 + \eta}\left[\frac{1}{q+i\delta} - \frac{1}{q-i\delta}\right] + \\
\frac{iE_{0}}{q+i\delta}\frac{\chi_+(q)}{\chi_+(0)}\frac{\varepsilon_{u-}(q)}{\varepsilon_{u-} (0)}\frac{\varepsilon_{g-}(0)}{\varepsilon_{g-}(q)}\frac{{e}^{-\chi(q)z_0}}{\varepsilon_u(q)}.
\end{multline}
The main ingredients of the solution are the effective dielectric functions of the 2DES in the ungated (u) and gated (g) sections that are given by
\begin{gather}
    \varepsilon_u (q)=1+i\eta \frac{\chi (q)}{{k}_{0}},\\
    \varepsilon_g (q)=1+i\eta \frac{\chi (q)}{{k}_{0}} (1-e^{-2\chi(q) d}),
\end{gather}
and $\chi(q)=\sqrt{q^2-k_0^2}$ is the decay constant of the electromagnetic field in the vertical direction. The gated and ungated dielectric functions have zeros at $q=\pm q_g$ and $q = \pm q_u$, which are the wave vectors of the gated and unagted plasmons in 2DES. To avoid ambiguities with the square root function, we set $k_0 = \omega/c + i\delta$, which corresponds to the tiny absorption in the ambient media. The branch cuts of the function $\chi(q)$ are chosen as rays running from $q = +k_0$ to $+i\infty$, and from $q = - k_0$ to $-i\infty$, respectively; therefore $\chi(q)$ is free of special points on the real axis. An important role in the solution is played by the factorized dielectric functions $\varepsilon_{g/u\pm}(q)$, such that 'plus' functions are analytic in the upper half-plane of the complex variable, and 'minus' functions are analytic in the lower half-plane. These functions are given by the Cauchy factorization formula:
\begin{equation}
\label{eq-factorisation}
{{\varepsilon}_{g/u \pm }}(q)=\exp \left\{ \pm \frac{1}{2i\pi }\int\limits_{-\infty }^{\infty }{\frac{\ln\varepsilon_{g/u} (u)du}{u-q\pm i0}} \right\}.
\end{equation}
The physical meaning of plus and minus functions appears clear after analyzing their zeros. The function $\varepsilon_{g-}$ has a zero at $q = + q_{g}$, and remains finite at $q = -q_g$. Similarly, the function $\varepsilon_+(q) \equiv \varepsilon(q)/\varepsilon_-(q)$ has a zero at $q = -q_u$ only. Therefore, the plus and minus functions carry information about the waves launched in the left and right half-spaces, respectively. It is now clear that the solution for electric field (\ref{eq-Eq_2d}) describes precisely the gated plasmons moving to the right ($q=+ q_g$) and the ungated plasmon moving to the left ($q=-q_u$). In agreement with physical intuition, other waves are not excited in the structure. 


\subsection{Theory of drag photovoltage}

The non-uniform local electric field in the 2DES drags the charge carriers. The emerging dc force ($\omega = 0$) is a result of local charge density accumulation in non-uniform electric field $\rho\propto \partial_x E_x$, and subsequent drag of this charge density by the $x$-component of the field. The resulting photovoltage ${V}_{\rm ph}$ can be presented both in real-space and Fourier-space representations~\cite{Popov_noncentrosymmetric,Svintsov_drag_contact,Durned_structured}:
\begin{multline}
		{{V}_{\rm ph}}=-\frac{2}{\omega Q n_{2d} }\int\limits_{-\infty}^{\infty}{{\rm Im}\left\{ {{\sigma }_{\omega }} E_{x}^{*} \frac{\partial E_x}{\partial x} \right\}dx} =\\
        +\frac{2{\rm Re}\sigma}{\omega Q n_{2d} }\int\limits_{0}^{\infty}{ \left[|E(q)|^2-|E(-q)|^2\right]\frac{q dq}{2\pi}}.	
\end{multline}
Above, $n_{2d}$ is the sheet density of charge carriers, $Q$ is the charge of the carrier ($-|e|$ for electron and $+|e|$ for hole), and the voltage is measured at the gated contact, such that the negative voltage corresponds to the force pushing the electrons under the gate.

To characterize sensitivity of the junction to the radiation, we introduce the photovoltage responsivity per incident light intensity $r_{\rm ph} = 8\pi V_{\rm ph} /E_0^2 \equiv V_{\rm ph}Z_0/(2E_0^2)$ where $Z_0 = 4\pi/c =377$ Ohm is the free-space impedance. This results in representation of responsivity as a product of dimensional prefactor (inversely proportional to frequency and carrier density) and a dimensionless {\it momentum transfer coefficient} $\alpha$
\begin{gather}
		\label{eq-rv-drag}
		r_{\rm ph} =  \frac{2}{\omega Q n_{\rm 2d}} \alpha,\\
        \label{alfa}
        {\alpha }=\text{Re} \eta
        \int\limits_{0}^{+\infty }{  \frac{\left| {E}\left( q \right) \right|^{2}-|{E}(q)|^{2}}{E_{0}^{2}} \frac{q dq}{2\pi }}.
\end{gather}
The above expression clarifies that the drag photovoltage is inversely proportional to the dimensional quantities $\omega$ and $n_{2d}$. The momentum transfer $\alpha$ is dimensionless, and is governed only by the dimensionless parameters characterizing the diffraction problem. There are two such parameters: the dimensionless 2DES conductivity $\eta$ and gate-channel separation normalized by the incident wavelength $k_0d = 2\pi d/\lambda_0$. Of course, $\eta$ can depend on frequency and carrier density by itself. Nevertheless, our solution of diffraction problem is insensitive to the microscopic origin of the conductivity $\eta$. The latter can be governed by intraband processes (Drude conductivity), interband radiation absorption, or both. We leave the precise determination of $\eta(n_{2d},\omega)$ to more detailed microscopic calculations~\cite{Falkovsky2007a} or spectroscopic experiments~\cite{Toksumakov2023}, and use it here as a free parameter.

\begin{figure}[ht!]
\centering
\includegraphics[width=0.9\linewidth]{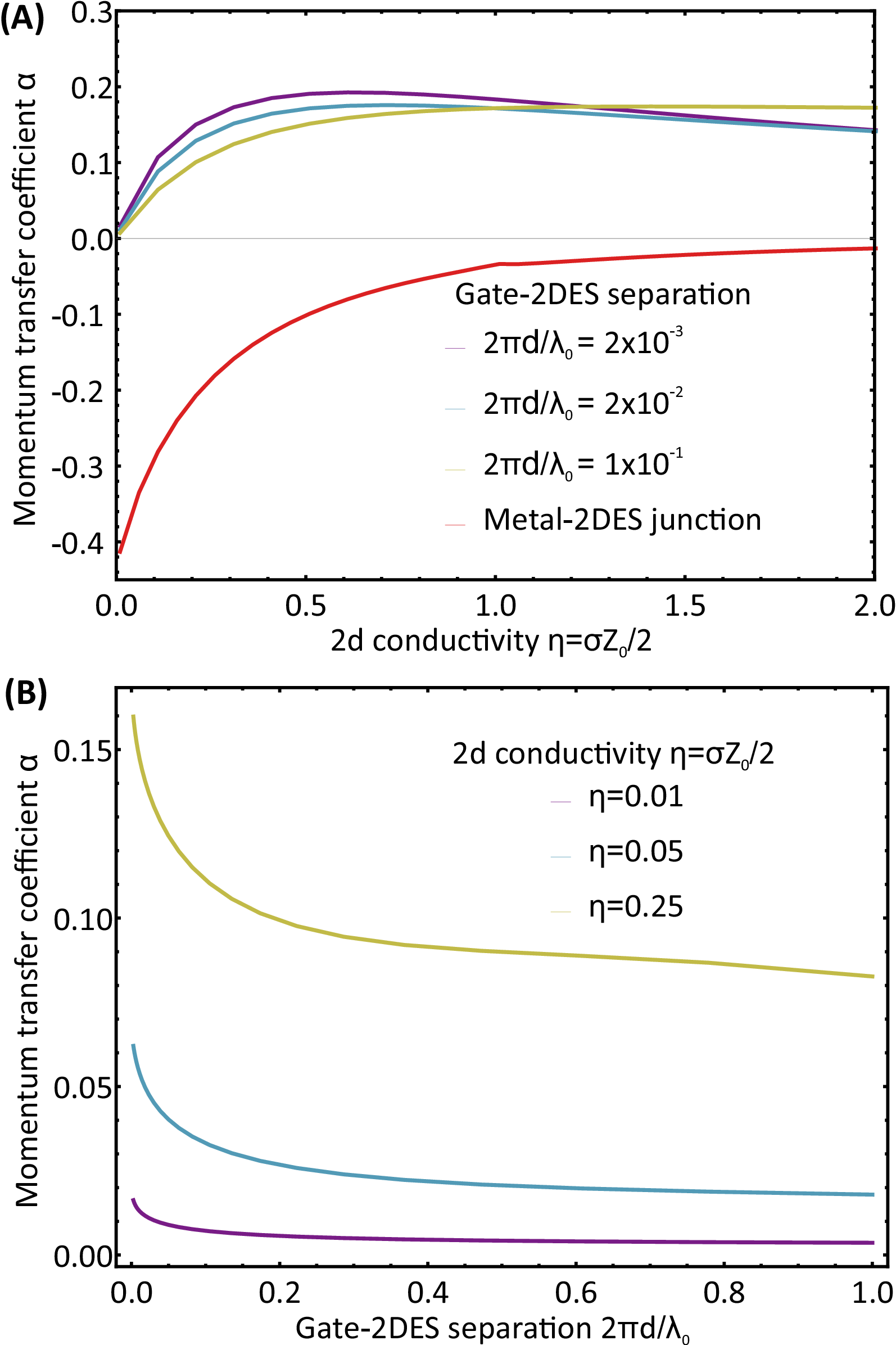}
\caption{Drag in partly gated 2DES for purely real conductivity. (A) Dependence of momentum transfer coefficient on dimensionless conductivity $\eta$ for various gate-2DES separations indicated in legend. Red curve shows the result for drag at the contact between metal and 2DES (B) Dependence of momentum transfer coefficient on gate-2DES separation $k_0d$ for selected values of $\eta$.}
\label{fig:drag_real}
\end{figure}

The study of photon drag voltage responsivity now reduces to the study of momentum transfer factor $\alpha$ as a function of $\eta$ and $k_0d$. We note that in most low-dimensional structures $k_0d\ll1$, and we shall further be interested in small values of $k_0d$ only. Indeed, the gate-channel separations lie between units of nanomenters and units of microns, while the wavelengths of detected radiation lie between millimeters (sub-THz range) to hundreds of nanomenters (visible range).

\subsection{Analysis of the drag responsivity}

We start the presentation of our results with the case of purely real conductivity $\eta$. The dependence of momentum transfer factor on real $\eta$ in shown in Fig.~\ref{fig:drag_real} (A) for $k_0d$ from $2\times10^{-3}$ to $10^{-1}$. First of all, we observe that $\alpha$ is order of unity, with maximum value $\sim 0.2$ reached for $\eta \sim 1$. It implies that photon drag in the partly gated setting is not small, and not limited by the value of in-plane electromagnetic momentum. The maximization of voltage responsivity at $\eta\sim 1$ has a purely electromagnetic origin: at smaller $\eta$ the 2DES has low intrinsic absorbance; at larger $\eta$ it acts as a mirror screening the external radiation. We also note that momentum transfer benefits from small gate-channel separations, at least for low values of $\eta$. This fact can be explained by the lighting-rod effect at the keen gate edge: the electric field at distance $d$ from the keen object can be estimated as $E_0(\lambda_0/d)^{1/2}$. The presence of 2DES violates this estimate only for large surface conductivity $\eta$, while at small $\eta$ the local field is continuously enhanced as $d$ scales down. The latter fact is illustrated in Fig.~\ref{fig:drag_real} (B), where the dependence of $\alpha$ on $k_0d$ is shown for selected conductivities.

The sign of momentum transfer coefficient is positive, which implies that the drag force pushes the charge carriers under the gate. In other words, the Fourier components of the electric field with positive $q$ are dominating over those with negative $q$. At this stage, it is instructive to compare $\alpha$ in the partly gated setting with that for metal-contacted 2DES $\alpha_0$. The latter is shown with red line in Fig.~\ref{fig:drag_real} (A). In both cases, the momentum transfer coefficient is order of unity. However, the signs of the force are different: the metal contact pushes the charge carriers away from 2DES.

\begin{figure}[ht!]
\centering
\includegraphics[width=0.9\linewidth]{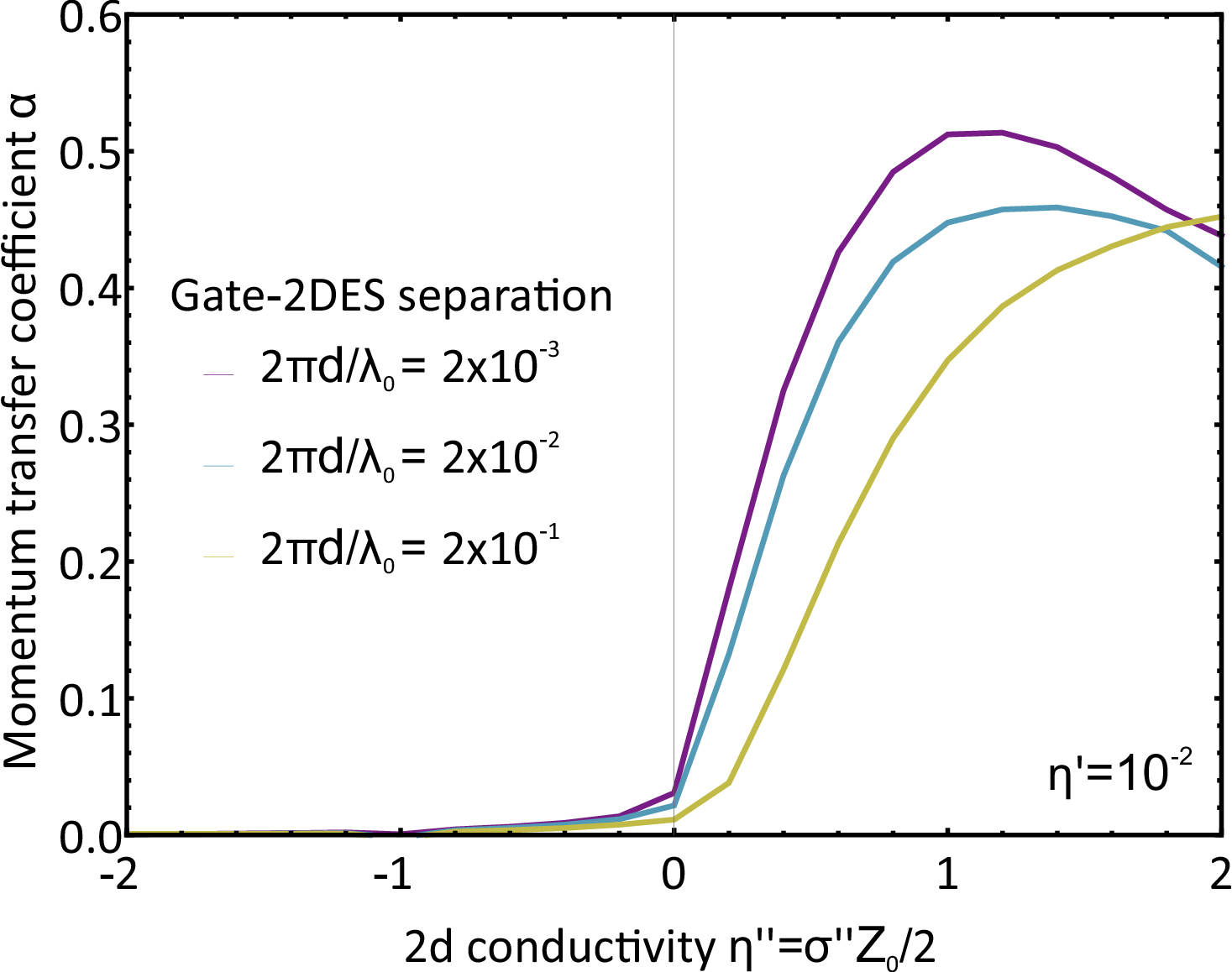}
\caption{Drag in partly gated 2DES for weakly dissipative 2DES conductivity. Dependence of momentum transfer coefficient on imaginary part of dimensionless conductivity $\eta''$ for various gate-2DES separations (indicated in legend). Real part of 2DES conductivity is taken constant and equal to $\eta\ = 10^{-2}$.}
\label{fig:drag_im}
\end{figure}

The origins of difference between drag force in partly gated and metal-contacted 2DES are becoming clearer after analyzing the case of reactive conductivity, $|\eta''| \gg \eta'$, where prime and double prime stand for real and imaginary parts of a complex quantity. The momentum transfer coefficient $\alpha$ vs imaginary part of conductivity is shown in Fig.~\ref{fig:drag_im}. Its remarkable property is large positive value for $\eta''>0$, and vanishingly small value for $\eta'' < 0$. The distinction between these two cases lies in the possibility of plasmon launching. The process is enabled for inductive-type conductivity $\eta''>0$, and disabled for capacitive-type conductivity $\eta'' < 0$~\cite{General_TE_Modes}. 

\begin{figure}[ht!]
\centering
\includegraphics[width=0.9\linewidth]{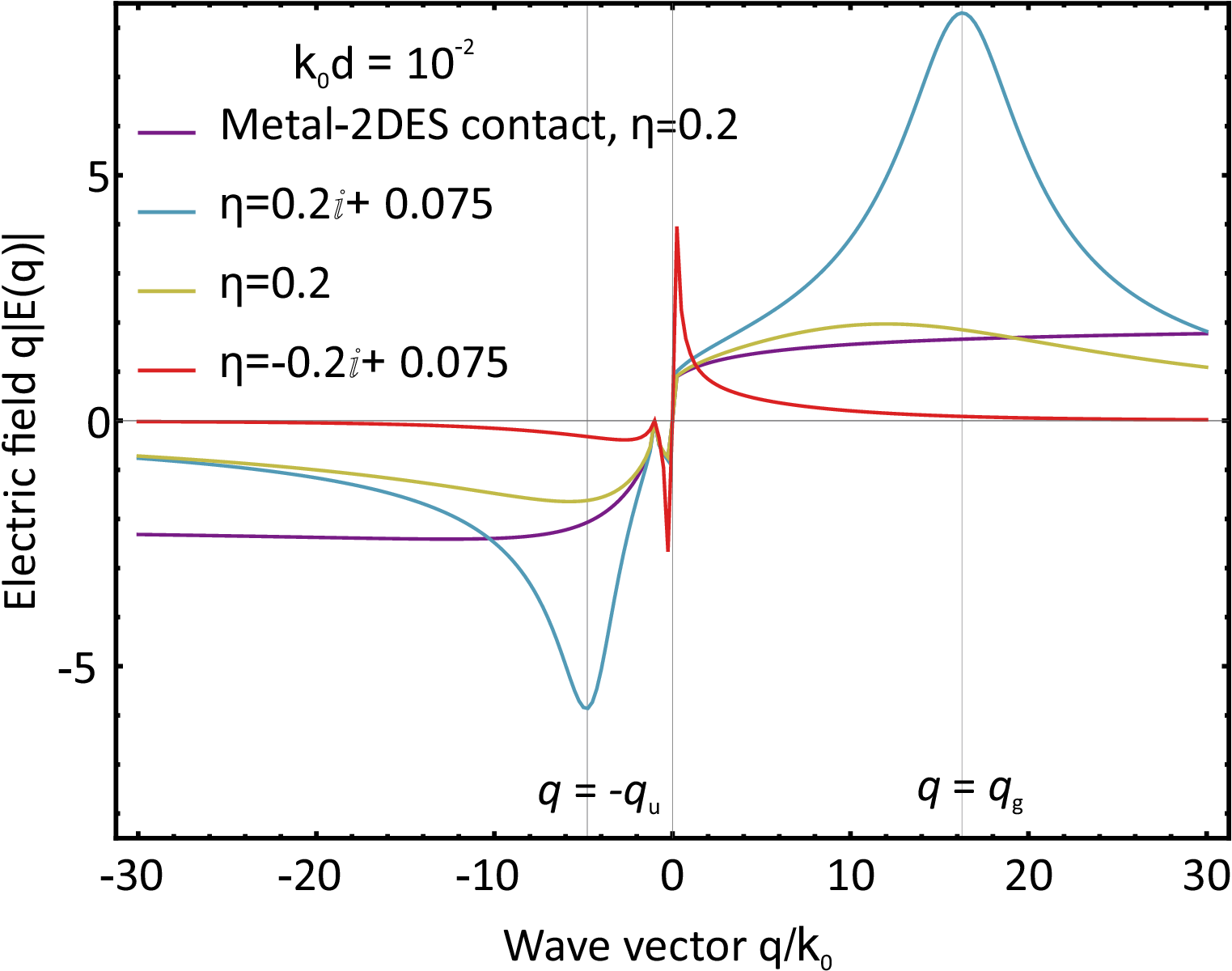}
\caption{Structure of the diffracted field and plasmonic origin of drag force. The dependence of Fourier spectrum of electric field in 2DES weighted with wave vector $q|E(q)|$ as a function of wave vector $q/k_0$ for several situations: metal-2DES contact (violet), partly gated 2DES with proximity gate ($k_0d=10^{-2}$) and inductive conductivity (blue), real conductivity (yellow), capacitive conductivity (red). In inductive case, the diffracted field develops resonances at plasma wave vectors $q=+q_g$ and $q=-q_u$. Some enhancement of field at $q\sim q_g$ persists for purely dissipative conductivity, while for capacitive conductivity the field is suppressed.}
\label{fig:efield_structure}
\end{figure}

The plasmonic origin of the drag force is fully confirmed by detailed inspection of the field profiles $E(q)$, which are shown in Fig.~\ref{fig:efield_structure} in the normalized form $q|E(q)|$. For inductive 2d conductivity (violet curve), the field spectrum is enhanced at $q=+q_g$ and $q=-q_u$. All the more, for purely real 2d conductivity (yellow curve), some enhancement of field Fourier harmonics is observed for positive wave vectors, as compared to the case of metal-2DES contact. Finally, field suppression is observed for almost all values of $q$ in the case of capacitive 2d conductivity. Therefore, launching of 2d plasmons by gate edge plays the dominant role in the formation of drag photovoltage.

The sign of drag force can now be analyzed in terms of plasmon amplitudes and momenta. More precisely, if the amplitudes of gated and ungated waves are $E_g$ and $E_u$, the sign of drag force would coincide with $q_g|E_g|^2-q_u|E_u|^2$. The gated plasmon wave vector always exceeds that of ungated one, $q_g>q_u$. In a practically interesting case of ultra-confined waves and proximized gates ($k_0d\ll |\eta| \ll 1$), these wave vectors are given by $q_u\approx i k_0/ \eta$ and $q_g \approx \sqrt{k_0/d\eta}$. All the more, analysis of field residues at the plasmon poles~\cite{Moiseenko_diffraction_launching} shows that the amplitude of gated wave is also larger, $|E_g|>|E_u|$. As a result, the drag force points from the ungated section to the gated one.  Similar arguments can be applied to the case of real conductivity, though the plasma waves are now overdamped. In that case, the ungated plasmon is purely evancescent, $q_g = i k_0/ |\eta|$. The gated plasmon, on the contrary, has both propagating and decaying components of the wave vector, $q_u \approx e^{i\pi/4}\sqrt{k_0/d|\eta|}$. The presence of finite propagation constant for gated mode explains the drag force pointing under the gate. The latter argument is also justified by the inspection of field spectra in Fig.~\ref{fig:efield_structure}: the positive Fourier components of the field are enhanced, while the negative components are suppressed in the partly gated setting, as compared to metal-2DES contact.

\begin{figure}[ht!]
\centering
\includegraphics[width=0.9\linewidth]{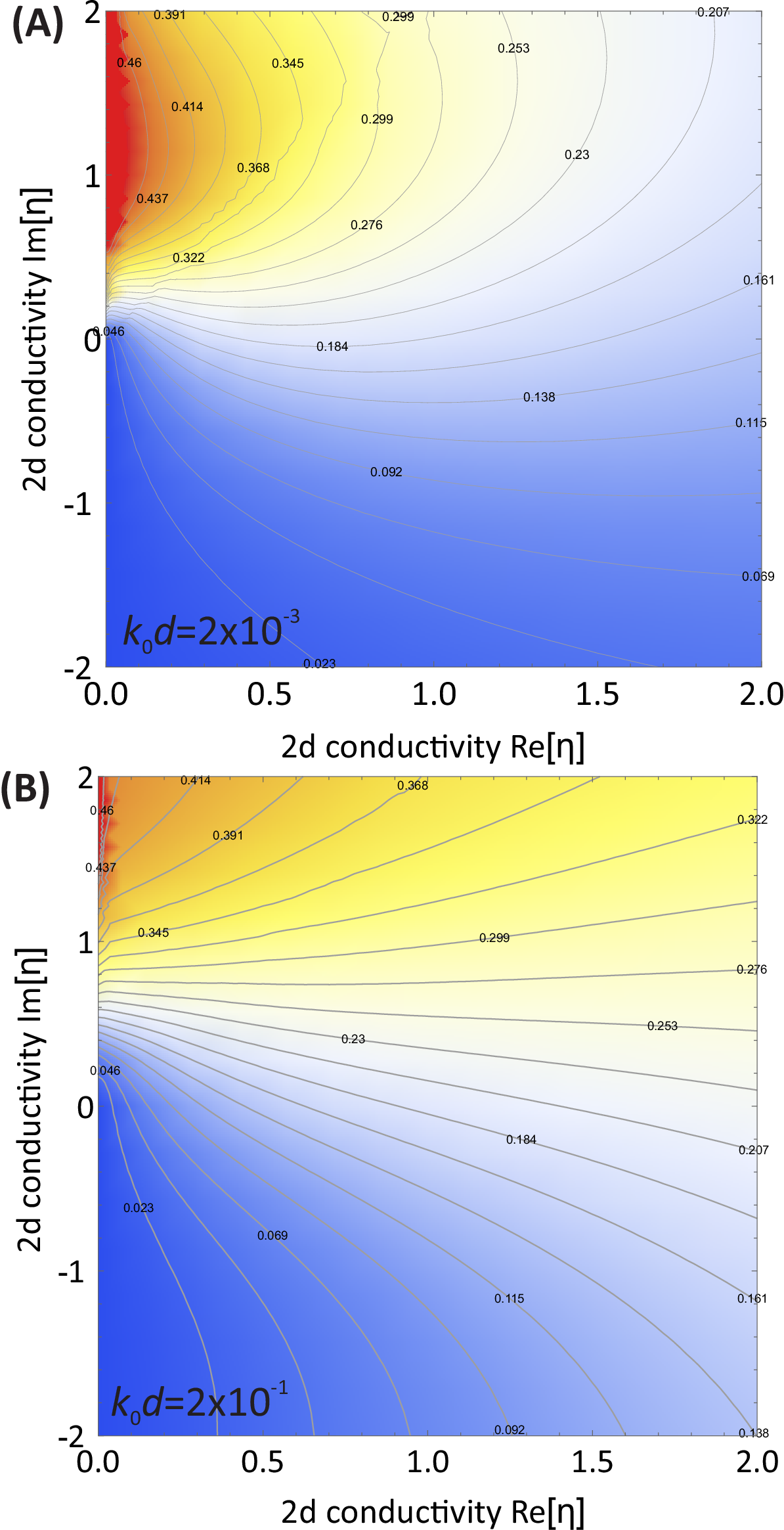}
\caption{Color maps of the momentum transfer coefficient $\alpha$ vs complex 2d conductivity $\eta'+i\eta''$ for small gate-2DES separation [panel (A), $k_0d=2\times10^{-3}$] and moderately large gate-2DES separation [panel (A), $k_0d=2\times10^{-2}$]. Hot colors correspond to large values of $\alpha$, cold colors -- to the values of $\alpha$ close to zero}
\label{fig:drag_complete}
\end{figure}

We conclude the presentation of our results with the values of momentum transfer coefficient $\alpha$ obtained for a broad range of complex-valued conductivities $\eta' + i \eta''$. The resulting dependences $\alpha(\eta',\eta'')$ are shown in Fig.~\ref{fig:drag_complete} as color maps. The maximization of $\alpha$ occurs at $\eta'' \sim 1$ and $\eta' \ll 1$. The maximization at large inductive conductivity is qualitatively explained by favorable conditions for the launching of 2d plasmons. At very large $\eta''$, the efficiency of plamson launching goes down due to the strong field screening by 2DES. The fact that momentum transfer does not depend on $\eta'$ is explained by its twofold role in the plasmon-assisted drag force. From one hand, small dissipation $\eta' \ll 1$ leads to the long free path of plasma wave. From the other hand, electromagnetic absorption is necessary for the transfer of momentum. Compensation of these two factors leads to an approximate independence of $\alpha$ on $\eta'$. As the gate-channel separation is increased [$k_0d=2\times10^{-1}$ in panel (B)], the conversion efficiency of external field into plasmons is reduced, and plasmonic maximum of responsivity is washed out.

\section{Discussion and conclusions}

We have revealed the properties of charge carrier drag by non-uniform fields in the basic block of most low-dimensional structures, the edge of the metal gate. The photovoltage developed due to the carrier drag does not contain the photon momentum as a small parameter; it appears large due to the efficient launching of gated 2d plasmons by the keen gate edge. The photovoltage responsivity is maximized for inductive-type 2DES conductivity with small dissipative part, these conditions are favorable for launching of 2d plasmons. For capacitive conductivity, the launching of plasmons is impossible, and the drag photovoltage appears very small.

The role of plasmons in the photovoltage generation in gated 2DES was pointed out previously in numerous papers on the so called Dyakonov-Shur rectification~\cite{Detection_mixing,Sakowicz2011}. Physically, it represents the same effect of electromagnetic drag due to the excitation of gated plasmons. We note, however, that the conventional Dyakonov-Shur approach postulates the solution for local electric field in the form of running plasma wave, and ignores all other non-propagating components of electric field. Our exact solution extends this approach to the situations where plasma waves are suppressed (i.e. capacitive type of 2d conductivity). Eventually, an exact solution allows us to link the amplitudes of launched plasmons to the incident field, i.e. to compute the conversion efficiency. Such estimate is challenging within the plane-wave matching approaches, where the modulation of external field by the gate appears as an external parameter~\cite{Gorbenko_LateralPC}.

The predictions of our model can be readily tested via measurements of photovoltage in partly gated 2DES. Exclusion of other mechanisms of rectification at the edge (e.g. thermoelectric and photovoltaic effects) is readily achieved by uniform 2DES doping. The latter is equivalent to holding the top gate at zero potential with respect the 2DES. The global back gate, if present, can control the overall carrier density in 2DES. In the case of graphene, tuning of Fermi energy $\mu$ can switch ${\rm Im}\eta$ from inductive type to capacitive type~\cite{Mikhailov_mode}. This provides an immediate opportunity to test the drastic reduction of drag photovoltage at ${\rm Im}\eta < 0$ predicted by our model.

The direct access to the drag photovoltage at an individual gate edge requires the suppression of the opposite edge. At moderately short wavelengths (visible to infrared), this can be achieved by radiation focusing~\cite{Gabor_thermoelectric}. At longer wavelengths (terahertz), this can be achieved by overlaying the gate and the drain contact. Instructively, the pulling of electrons by the incident wave under the gate can lead to measurable effects even in the symmetric gated structures. Namely, electron accumulation under the gate would lead to the shift of resistance-gate voltage curve $\rho(V_g)$ precisely by $V_{\rm ph}$. The shift attains readily measurable values. Particularly, focusing the power $P=10$ mW (typical for solid-state sub-THz diodes) into the area of 10 mm$^2$, taking the frequency $f=100$ GHz, carrier density $n_{2d}=10^{11}$ cm$^{-2}$ and momentum transfer factor $\alpha=0.2$, we get $V_{\rm ph}\approx 4$ $\mu$V. Accurate estimates of the local fields generated upon diffraction at finite-size gates can be obtained with the modified local capacitance approximation~\cite{Zabolotnykh_proximity,Enaldiev_Near_Gate}.

A natural limitation of our exact solution for local electric field and, hence, electromagnetic drag, is the uniformity of 2DES conductivity. The case of conductivity steps (possibly formed upon gate voltage application) is currently beyond the capability of Wiener-Hopf method. A reasonable analytical alternative to it, especially for finite-size gates, lies in the use of trial functions for current distributions in the gates~\cite{Mikhailov1998,Khisameeva_plasmonic_crystal}. The method is appealing for analytical analysis of drag in the complex multi-gate structures~\cite{Popov_noncentrosymmetric,Olbrich2016}.

\section{Funding}
The work was financially supported by the grant \# 21-79-20225 - P of the Russian Science Foundation. 	
    
\bibliography{Bibliography}
	
\end{document}